\documentstyle[aps]{revtex}
\draft
\begin{document}
\title{Two-Dimensional Nature of Four-Layer Superconductors 
       by Inequivalent Hole Distribution}
\author{Mun-Seog Kim, C. U. Jung, and Sung-Ik Lee}
\address{National Creative Research Initiative Center for Superconductivity and
         Department of Physics, Pohang University of Science and Technology,
         Pohang 790-784, Republic of Korea}
\author{A. Iyo}
\address{Electrotechnical Laboratory, Umezono 1-1-4, Tsukuba, Ibaraki 305, Japan}
\maketitle

\begin{abstract}
The magnetization of the four-layer superconductor 
CuBa$_{2}$Ca$_{3}$Cu$_4$O$_{12-\delta}$ with $T_c\simeq117$ K is presented.
The high-field magnetization around $T_c(H)$ follows the exact two-dimensional 
scaling function given by Te\v{s}anovi\'{c} and Andreev. This feature is 
contrary to the inference that the interlayer coupling becomes strong 
if the number of CuO$_2$ planes in a unit cell increases.
Also, the fluctuation-induced susceptibility in the low-field region was analyzed 
by using the modified Lawrence-Doniach model. The effective number of 
independently fluctuating CuO$_2$ layers per unit cell, $g_{\rm eff}$, 
turned out to be $\simeq2$ rather than 4, which indicated that two among the 
four CuO$_2$ layers were in states far from their optimal doping levels. 
This result could explain why CuBa$_{2}$Ca$_{3}$Cu$_4$O$_{12-\delta}$ 
shows two-dimensional behavior.

\end{abstract}
Within a high-$T_c$ homologous series, the $T_c$ is expected to increase with the 
number of CuO$_2$ planes per unit cell, $n$, because an increase in $n$ means an 
increase in the number of CuO$_2$ planes per unit volume and thereby 
increases in the charge-carrier density and the coupling between the conducting planes.
In fact, the $T_c$ increases with $n$ within a high-$T_c$ family. However, this 
tendency does not persist above a certain value of $n$. 
For example, within the HgBa$_2$Ca$_{n-1}$Cu$_{n}$O$_{2n+2+\delta}$ family, 
the $T_c$ increases with $n$ up to $n=3$, but for $n=4$ the value is lower by 
about 10 K in comparison with the value for $n=3$. 

For compounds with $n\geq 3$, the unit cells contain two structurally different 
CuO$_2$ planes. In the case of HgBa$_2$Ca$_3$Cu$_4$O$_{10+\delta}$ (Hg-1234),
among the four layers, two layers contain Cu with square-planar coordination (Type-A), 
and the other two contain Cu with pyramidal coordination (Type-B). 
Yamauchi and Karppinen\cite{yamauchi1} claimed from their bond-valence-sum (BVS) 
calculations that the charge carriers can be inhomogeneously distributed 
between Type-A and Type-B CuO$_2$ planes and that the holes may be concentrated 
mainly in the Type-A CuO$_2$ planes. 

As a consequence of the inequivalent hole distribution, the ``{\em microscopic}'' 
$T_c$'s of the  Type-A and Type-B planes will differ from each other.\cite{memo2}
The difference between the lower and the higher $T_c$'s becomes severe as the 
degree of imbalance in the hole distribution increases. Thus, at temperatures 
around the higher $T_c$, one of the two different kinds 
of CuO$_2$ planes does not play the role of a superconducting layer by itself; 
hence the interlayer spacing in the system can be effectively quite large. 

In our previous works,\cite{mskim14,mskim17} we demonstrated that in high-field region,
the thermal fluctuations of Hg-1234 show two-dimensional (2D) scaling behavior 
around $T_c(H)$. Furthermore, the strong anisotropic nature of this compound 
as observed through magnetic torque measurements by Zech {\em et al.}\cite{zech1} 
They reported the anisotropy ratio, $\gamma=\lambda_c/\lambda_{ab}$, 
of Hg-1234 to be about 52. These results implying a weak interlayer coupling are 
consistent with our viewpoint that the inequivalent distribution of holes can
effectively cause a large interlayer spacing. 

The above studies prompted an examination of whether or not an inequivalent 
distribution of holes is a common feature in four-layer cuprates. 
With this aim, we measured the reversible and fluctuation-induced 
magnetization of another four-layer system 
CuBa$_2$Ca$_3$Cu$_4$O$_{12-\delta}$\cite{ihara1,ihara2,alario1,gao1,wu1} (Cu-1234) 
with $T_c=117$ K. The main difference between Cu-1234 and Hg-1234 lies in 
the structure of the charge reservoir block (CRB). While the CRB of Hg-1234 
consists of a double rock-salt block [BaO][HgO$_\delta$][BaO] $(\delta < 1)$, 
Cu-1234 contains a CuO$_{2-\delta}$ plane instead of a HgO$_\delta$ plane 
in the block. The $c$-axis parameter of Cu-1234 is shorter than that of 
Hg-1234 by about 1 \AA~due to the relatively thin CRB. 
The $T_c$, 117 K, of Cu-1234 is known not to vary even after post 
annealings under various conditions.\cite{ito2} 

In this letter, our intention is to elucidate the dimensional nature and 
the superconducting properties of Cu-1234. The measured magnetization 
data were analyzed using the high-field scaling theory 
proposed by Te\v{s}anovi\'{c} and Andreev,\cite{tesanovic1} 
the modified Lawrence-Doniach model,\cite{lawrence1,klemm1,prober1} 
and the Hao-Clem model.\cite{hao1,hao2,mskim6,mskim9,mskim15} 
From these analyses, we found that Cu-1234 had a strong 2D nature 
which was caused by an effective reduction of the 
number of CuO$_2$ planes due to an inequivalent hole distribution. 

The details on the sample preparation under a high-pressure 
condition ($\sim$ 4 Gpa) are given elsewhere.\cite{iyo1} 
The lattice parameters, $a=b=3.858$ \AA~and  $c=17.98$ \AA, 
were obtained from X-ray diffraction. To obtain a $c$-axis aligned sample, 
we employed the Farrell method.\cite{farrell1} First, we passed the powder 
of the sample through a 20-$\mu$m sieve to filter out grains with 
multi-domains. This fine powder was aligned in commercial 
epoxy (Hardman, Inc.) under an external magnetic field of 7 T. 
The dimension of the permanently aligned sample was approximately
9.5 mm in length and 3 mm in diameter. From the X-ray 
rocking-curve measurement, the full width at half maximum (FWHM) 
of the (006) reflection was found to be less than 2 degrees.
The temperature dependence of the magnetization was measured
in the magnetic field range of 0.05 T $\leq H \leq$ 5 T by using
a SQUID magnetometer (MPMS, Quantum Design).

Various thermodynamic parameters characterizing Cu-1234 were evaluated by applying
the Hao-Clem model to the reversible magnetization measured in the field 
range 1 T $\leq H \leq$  5 T. Figure\ \ref{fig1} shows the temperature 
dependence of the thermodynamic critical field $H_c(T)$ obtained
from the above analysis. In this figure, the solid line represents 
the BCS temperature dependence of $H_c$.\cite{clem1} This model 
yields $H_c(0)=0.9$ T and $T_c=117$ K, which corresponds to a slope 
of $dH_c/dT=-129$ Oe/K near $T_c$. Using the relationship
$H_{c2}(T)=\sqrt{2}\kappa H_c(T)$ and employing $\kappa=127$, 
which was deduced from the Hao-Clem model, 
we estimated the upper critical field slope 
as $(dH_{c2} / dT)_{T_{c}}=-2.3$ T/K. This slope can be used
to calculate the upper critical field at zero temperature by using 
the Werthamer-Helfand-Hohenberg formula,\cite{werthamer1} 
and $H_{c2}(0)$ was estimated to be 196 T [$\xi_{ab}(0)=12.8$ \AA] 
in the clean limit. The penetration depth $\lambda(T)$ was evaluated 
by using the relationship 
$\lambda = \kappa(\phi_0/2\pi H_{c2})^{1/2}$, 
as shown in inset of Fig.\ \ref{fig1}. The solid line in the inset represents 
the penetration depth $\lambda_{ab}(T)$ in the clean limit 
with $\lambda_{ab}(0)=198$ nm. In Table\ \ref{tab1}, all these parameters
are summarized along with those of Hg-1234\cite{mskim14} for comparison. 
With this preliminary information, we now proceed to study 
other superconducting properties of Cu-1234.

Figure\ \ref{fig2} shows the irreversibility line of Cu-1234 (open circles) 
obtained from the DC magnetization
curves $4\pi M(T)$ for 0.1 T $\leq H \leq$ 5 T. The open squares in the
figure denote the data for Hg-1234. We note that the irreversibility line 
of Cu-1234 is shifted to higher temperature in comparison with that of Hg-1234. 
This implies that the vortex pinning in Cu-1234 is more effective. 
It is generally accepted that a strong interlayer coupling gives rise 
to a strong flux pinning. As mentioned in the introduction, the interlayer 
spacing of Cu-1234 is smaller than that of Hg-1234.
We conjecture that this causes an enhanced interlayer coupling. 

Because the interlayer coupling of Cu-1234 is rather strong, one can expect 
a more enhanced superconductivity. However, the transition temperature of Cu-1234 
is lower than  the value of 125 K for Hg-1234 by 8 K. Not only the interlayer coupling
strength\cite{wheatley1,chakravarty1,tesanovic3,ye1} but also the carrier
concentration is known to be responsible for determining the transition
temperature of layered superconductors.\cite{cooper1} 
Thus, it is easily postulated that the relatively low $T_c$ of Cu-1234 
compared to that of Hg-1234 
is due to a low carrier density within the conducting planes. 
As shown in Table\ \ref{tab1}, the penetration depth of Cu-1234, 
$\lambda_{ab}(0)=198$ nm, is considerably larger than that of Hg-1234.
If we assume that the electronic effective mass in the $ab$ plane, 
$m^\ast_{ab}$, is nearly the same as  that of Hg-1234, this larger value 
of $\lambda_{ab}(0)$ justifies the above postulation 
through the relationship $\lambda_{ab}(0)\propto(m^\ast_{ab}/n_s)^{1/2}$.

Previously, we reported that Hg-1234 is a strong 2D 
superconductor.\cite{mskim14,mskim17} The direct evidence for this was 
based on a scaling analysis of the fluctuation-induced 
magnetization in the high-field region. The same analysis using the 
high-field scaling function was applied to Cu-1234. 
In the high-field limit, according to 
Te\v{s}anovi\'{c} and Andreev,\cite{tesanovic1}
the exact scaling function for a 2D system is given by
        \begin{equation}
        \frac{M(H,T)}{\sqrt{HT}}\frac{\phi_0 H_{c2}^\prime d}{A}=
        x-\sqrt{x^2+2}\label{tesa2d},
         \end{equation}
where $A$ is a constant, $x=A[T-T_c(H)]/(TH)^{1/2}$,
$H_{c2}^\prime=(dH_{c2}/dT)_{T_c}$, and
$d$ is the effective interlayer spacing.
To compare the theory with our data, we used a modified form of Eq.\ (\ref{tesa2d}): 
        \begin{equation}
        \frac{M}{M^\ast}= \frac{1}{2}\{1-\tau-h+\sqrt{(1-\tau-h)^2+4h}
        \}\label{tesa2dr},
         \end{equation}
where $M^\ast$ is the field-independent magnetization which is observed at a
certain temperature $T^\ast(<T_c)$, $\tau=(T-T^\ast)/(T_c-T^\ast)$, 
and $h=H/H_{c2}(T^\ast)$. Figure\ \ref{fig3} shows our attempt to fit 
the fluctuation-induced magnetization data by using Eq.\ (\ref{tesa2dr}) with
the experimental values of $M^\ast=-1.8$ G and $T^\ast=114$ K.
The scaled magnetization curves for various values of the field are
shown in the inset of Fig.\ \ref{fig3}. The solid lines represent the 
theoretical curves. This analysis gives $T_c=117$ K and 
$(dH_{c2}/dT)_{T_c}=-2.27$ T/K, which are
fairly consistent with the results from the Hao-Clem analysis.
On the other hand, the fit using the 3D version of the scaling function proposed by
Ullah and Dorsey\cite{ullah1} was less satisfactory.
As stated before, the interlayer coupling of Cu-1234 is enhanced
compared with that of Hg-1234. However, the above scaling result indicates that
in spite of the smaller interlayer distance, the coupling strength of Cu-1234
is still weaker than those of 3D superconductors such as 
YBa$_2$Cu$_3$O$_y$ and YBa$_2$Cu$_4$O$_y$.

Finally, we measured the temperature dependence of the 
fluctuation-induced magnetic susceptibility
at fields of 500 and 1000 Oe, as shown in Fig.\ \ref{fig4}.
In the modified Lawrence-Doniach model,
the fluctuation-induced diamagnetic
susceptibility in a 2D system is given by

    \begin{equation}
    \chi_c^{2D}(T)=-g_{\rm eff} \frac{\pi k_B T \xi_{ab}^2(0)}{3\phi_0^2s} \bigg( \frac{T_c}{T-T_c}\bigg),
    \label{lf2d}
    \end{equation}
where $s$ is the $c$-axis repeat distance and $g_{\rm eff}$ is the effective number of 
independently fluctuating CuO$_2$ layers per unit cell.

The solid lines of Fig.\ \ref{fig4} represent least-squares fits of Eq.\ (\ref{lf2d})
in the temperature range of $T>T_c$.
From this analysis, we obtain $g_{\rm eff}\pi T_c k_B \xi_{ab}^2 / 3 \phi_0^2 s$
= 4.7 $\times 10^{-8}$ and $T_c=116.4$ K. 
If we employ $s=17.9$ \AA~and $g_{\rm eff}=4$,
then the coherence length $\xi_{ab}(0)$ is estimated to be 7.3 \AA.
However, this value is significantly smaller than 
the value of $\xi_{ab}(0)=12.8$ \AA~ obtained from the
Hao-Clem model and the high-field scaling analyses.
This discrepancy strongly implies that the value of $g_{\rm eff}$ is less than four.
For comparison, we reexamined the temperature dependence of the fluctuation-induced
magnetic susceptibility of Hg-1234, which is also shown in Fig.\ \ref{fig4}.
The $\xi_{ab}(0)$ for Hg-1234 is estimated to be 9.6 \AA~based on 
the modified Lawrence-Doniach model. This value is also smaller than the value of 
$\xi_{ab}(0)=12.7$ \AA~presented in Table\ \ref{tab1}.

A possible scenario to explain these experimental results is as follows: 
For Cu-1234 and Hg-1234, among the four conduction layers, two CuO$_2$ layers 
are bridged to the charge reservoir block by apical oxygen.
However, the other two layers have an infinite-layer structure without apical oxygen. 
This structural feature might cause the imbalance in the charge-carrier concentration 
between the two different kind of CuO$_2$ planes, as suggested by 
the BVS calculation.\cite{yamauchi1} If one of the two kinds 
of CuO$_2$ layers is in a strongly overdoped (or underdoped) state, 
the superconductivity in the layers could be highly suppressed.
In this context, one can assume $g_{\rm eff}=2$ rather than 4. 
Assuming $g_{\rm eff}=2$, we recalculated the $\xi_{ab}(0)$'s and obtained 
10.3 \AA~and 12.1 \AA~for Cu-1234 and Hg-1234, respectively.
Compared with the values obtained assuming $g_{\rm eff}=4$,
these values are close to the values obtained from the Hao-Clem analysis

In a high-$T_c$ homologous series, 
the transition temperature increases until a certain value of $n$ and then 
slowly decreases for higher values of $n$. The Cu-based homologous series 
shows the same feature. As in the Hg-based series, the compound with $n=3$ has the
maximum $T_c$.\cite{ito2,ito1} From the above analysis
of the magnetic susceptibility, we can infer that such a decrease in $T_c$
with $n$ for high-$T_c$ compounds of $n \geq 4$
might be due to the CuO$_2$ planes that do not play roles as superconducting layers.

In summary, the magnetization $4\pi M(T)$ of $c$-axis oriented Cu-1234
was measured in the field range of 0.05 T $\leq H \leq$ 5 T.
In comparison with Hg-1234, the irreversible region in the $H$-$T$ phase diagram
is more broader, which originates from enhanced interlayer coupling due to the 
relatively short $c$-axis parameter.
However, from the high-field scaling analysis of the magnetization around $T_c(H)$,
the dimensionality of Cu-1234 is found to be still two dimensional.
Our experimental results for the magnetic susceptibilities of 
four-layer compounds (Cu-1234 and Hg-1234) suggest that two among the four 
CuO$_2$ layers do not contribute to the superconductivity 
due to an inequivalent hole distribution between the two different CuO$_2$ planes. 
This could explain the origin of the weak interlayer coupling in four-layer 
superconductors and provide the reason for the $T_c$ of compounds with 
$n \geq 4$ decreasing with $n$. In other words, if the optimum number of 
holes is doped into all the CuO$_2$ planes equivalently,  stronger 
interlayer coupling, and thereby a higher $T_c$, can be achieved.

We thank M. Sigrist, D. Pavuna, M. Salamon, F. Borsa, and Y. Mori
for useful discussions. This work was supported by the Creative Research 
Initiatives of the Korean Ministry of Science and Technology.

\begin{table}
\caption{Thermodynamic parameters of CuBa$_{2}$Ca$_{3}$Cu$_4$O$_{12-\delta}$ and
         HgBa$_{2}$Ca$_{3}$Cu$_4$O$_{10+\delta}$ superconductors 
         deduced from the reversible magnetization.}
\begin{tabular}{lcc}
&Cu-1234&Hg-1234\\ \hline
$T_c$ (K)                       & 117                   & 125\\
$\kappa$                        & 127                   & 102\\
$-(dH_{c2}/dT)_{T_c}$ (T/K)     & 2.3                   & 2.2\\
$H_c(0)$ (T)                    & 0.9                   & 1.1\\
$H_{c2}(0)$ (T)                 & 196                   & 205\\
$\xi_{ab}(0)$ (\AA)             & 12.8                  & 12.7\\
$\lambda_{ab}(0)$ (nm)          & 198                   & 157
\end{tabular}
\label{tab1}
\end{table}

\begin{figure}
\caption{Temperature dependence of the thermodynamic critical field $H_c(T)$
         extracted by using the Hao-Clem model. The solid line represents the
         BCS temperature dependence of $H_c(T)$. The inset shows the 
         temperature dependence of the penetration depth 
         $\lambda_{ab}(T)$, and the theoretical curve (solid line)
         assumes the BCS clean limit.}
\label{fig1}
\end{figure}

\begin{figure}
\caption{Irreversibility lines of CuBa$_{2}$Ca$_{3}$Cu$_4$O$_{12-\delta}$ and
         HgBa$_{2}$Ca$_{3}$Cu$_4$O$_{10+\delta}$.}
 \label{fig2}
\end{figure}

\begin{figure}
\caption{Temperature dependence of the magnetization $4\pi M(T)$
         around $T_c$. The solid lines represent theoretical curves obtained 
         by using the exact scaling function proposed by Te\v{s}anovi\'{c} and Andreev 
         ($\circ~1$ T, $\Box~2$ T, $\bigtriangleup~3$ T, $\bigtriangledown~4$ T, and
         $\Diamond~5$ T). The inset shows 2D scaling of
         the fluctuation-induced magnetization 4$\pi M(T,H)$.}
         \label{fig3}
\end{figure}
\begin{figure}
\caption{Temperature dependence of the fluctuation-induced susceptibility
         $\chi (T)$ for Cu-1234 and Hg-1234. The solid lines represent
         the modified Lawrence-Doniach model for a 2D system.}
\label{fig4}
\end{figure}


\begin{references}
\bibitem{memo2}
However, in measurements of macroscopic quantities, such as the magnetization
  or the resistivity, we can only observe the superconducting transition of the 
  CuO$_2$ planes having higher $T_c$.
\end{references}
\end{document}